\documentclass[12pt]{article}
\usepackage{epsfig}

\newcommand{\ie}{{\it i.e. }} 
 
\newcommand{\beq}{\begin{equation}} 
\newcommand{\eeq}{\end{equation}} 
\newcommand{\beqa}{\begin{eqnarray}} 
\newcommand{\eeqa}{\end{eqnarray}}

\begin{document}
\thispagestyle{empty} 
\begin{flushright} 
physics/0112056 
\end{flushright} 
\vskip 3cm 
\begin{center} 
{\Large \bf Linearisation of Simple Pendulum} 
\vskip 1cm 
{\sf \bf P. Arun\footnote{arunp92@yahoo.co.in} \& 
Naveen Gaur\footnote{naveen@physics.du.ac.in,~pgaur@ndf.vsnl.net.in }} \\ 
{\sf Department of Physics \& Astrophysics} \\ 
{\sf University of Delhi}\\ 
{\sf Delhi - 110 007, India} 
\end{center}
\vfill 
\begin{abstract} 
The motion of a pendulum is described as Simple Harmonic Motion (SHM) in case 
the initial displacement given is small. If we relax this condition then we 
observe the deviation from the SHM. The equation of motion is non-linear and 
thus difficult to explain to under-graduate students. This manuscript tries 
to simplify things. 
\end{abstract}
\vfill 
\pagebreak 
One of the basic experiments which a physics student will try out is 
that of the pendulum. A pendulum consists of a massive bob 
suspended from a massless string, in actual terms, the string does have 
a mass but it is negligible as compared to the mass of the bob. Also, for 
making the assumption that the bob is a point mass, the length of the string 
is made far greater than the radius of the bob.
\par The general equation of motion (EOM) of pendulum was derived in 
our previous work \cite{book1,parun1}. The equation \cite{book1} was : 
\beq 
cos{\Theta \over 2} ~ \frac{d^2\Theta}{dt^2} ~-~ {1 \over 2} 
sin{\Theta \over 2}(\frac{d\Theta}{dt})^2 ~=~ - ~\omega^2 ~sin 
\Theta 
\label{eq:1} 
\eeq 
In above eq.(\ref{eq:1}) $\Theta$ is the angular displacement and 
$\omega = \sqrt{\frac{g}{l}}$, where the terms have their usual meaning. 
This is a non-linear differential equation and is not solvable 
analytically. We adopted numerical methods to get the solutions of this 
equation under various conditions. If the initial displacement is
small, \ie  under small oscillation approximation we can 
write $sin \theta \sim \theta$ and $cos \theta \sim 1$. Using this, 
eq.(\ref{eq:1}) goes to 
\beq 
\frac{d^2\Theta}{dt^2} - {1 \over 4} \Theta (\frac{d\Theta}{dt})^2 
~=~ - \omega^2 \Theta 
\label{eq:2} 
\eeq 
the second term on the left side of above eqn. is a second order term 
in $\Theta$, so we can neglect this term being the higher order 
term (in $\Theta$) . So the eqn becomes 
\beq 
\frac{d^2\Theta}{dt^2} ~=~ - \omega^2 \Theta 
\label{eq:3} 
\eeq 
this is the equation of SHM. So under small oscillation approximation 
the EOM of non-linear pendulum turns to a linear second order differential 
equation in time, more well known as the simple harmonic motion. 

\par Let's now discuss the solutions of both linear (eq.(\ref{eq:3}))
and non-linear (eq.(\ref{eq:1})) equations. As we discussed earlier
work \cite{parun1} that the eq.(\ref{eq:1}) can't be solved
analytically, so we tried out numerical solution for the non-linear
equation. But the solutions of linear equation(\ref{eq:3}) are well
known and can be written as :
\beq 
\Theta_i(t) ~=~ A Cos(\omega_i t) + B Sin(\omega_i t) 
\label{eq:4} 
\eeq 
where A and B are the constants (we expects two constants to be there 
in the general solution as the equation which we are trying to solve 
is a second order equation in time), and $\omega_i = \sqrt{2 \pi \over 
T_i}$, where the i subscript indicates that the EOM is SHM \ie a
linear second order equation. The 
solution (equation \ref{eq:4}) is a harmonic function of 
time. One can determine the unknown coefficients A and B by using the 
initial conditions. 

\begin{figure}[htb]
\begin{center}
\epsfig{file=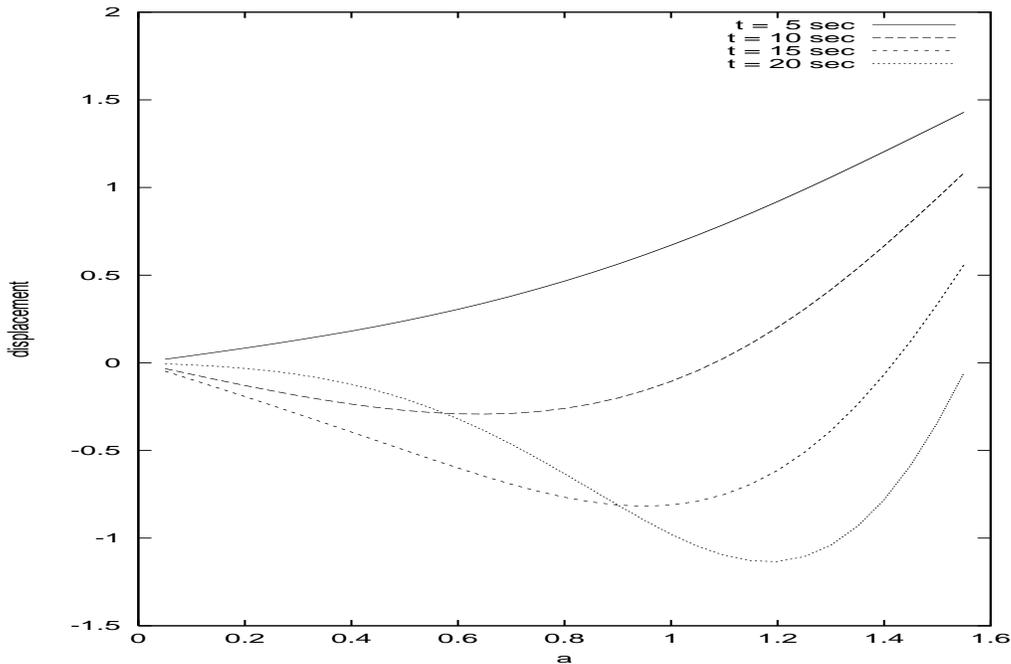,width=5.5in,height=3.5in}
\caption{The plot of the displacement at various times (t = 5, 10, 15,
20 seconds) with the initial displacement, $\omega = {\rm 4}$ }
\end{center}
\label{fig:1}
\end{figure}

\par For a simple pendulum the total time taken for completing N
oscillations can be written as :
\beq
t ~=~ T_1 + T_2 + \dots + T_N
\label{eq:5}
\eeq
where $T_1$, $T_2$, \dots $T_N$ are time periods of first, second,
\dots and $N^{th}$ oscillation respectively. For a simple pendulum
undergoing SHM all the time periods are same \ie
\beq
T_1 ~=~ T_2 ~=~ \dots ~=~ T_N ~=~ T_i
\label{eq:6}
\eeq
leading to
\beq
t_i ~=~ N T_i
\label{eq:7}
\eeq
where the subscript i indicates the SHM. However in the case of
pendulum obeying the non-linear EOM (eq.(\ref{eq:1}), as we have shown
in previous work \cite{parun1}, the time period of oscillation won't
be the same as in the case of SHM (in fact the displacement given by
non-linear EOM lags behind in phase to that of SHM). Let's
parameterise this lag in phase by saying that a constant phase
difference (say $\alpha$) is introduced to the time period after each
successive oscillation \ie
\beq
T ~=~ T_i - \alpha
\label{eq:8}
\eeq
where T is the time period of the pendulum using non-linear EOM. So
after N oscillations the difference between the two times would be 
\beq 
t ~=~ t_i - N \alpha 
\label{eq:9} 
\eeq 
where ${\rm t}$ and ${\rm t_i}$ are the time taken to complete N
oscillations by the non-linear and linear oscillators respectively,
while $\alpha$ is the small variation introduced with each successive
oscillations. As time increases, the pendulum whose motion is
described by eq(\ref{eq:1}) completes more oscillations as compared to
the simple pendulum, for a given time. Thus, the only correction
called for seems to be a correction factor in the angular frequency.
As we have also shown earlier \cite{parun1} that the difference
between the two time periods increases as the initial amplitude of the
oscillation increases. So we can say that the correction factor is
dependent of the initial displacement. Also, for the initial
displacement tending to be small, the correction factor should
approach zero and hence the non-linear oscillator should approach the
SHM. So effectively we can say that the motion of non-linear
oscillator can still the thought of simple harmonic, with the
difference that the frequency is now dependent on the initial
amplitude. So we can write solution of actual pendulum of type :
\beq 
\Theta(t) ~=~ a ~Cos( \omega_i t + f(a) t) 
\label{eq:10}
\eeq 
where the constant A has been replaced with initial amplitude a. The 
function ${\rm f(a)}$ is the function of initial displacement as
argued above . 
\begin{figure}[htb] 
\begin{center} 
\epsfig{file=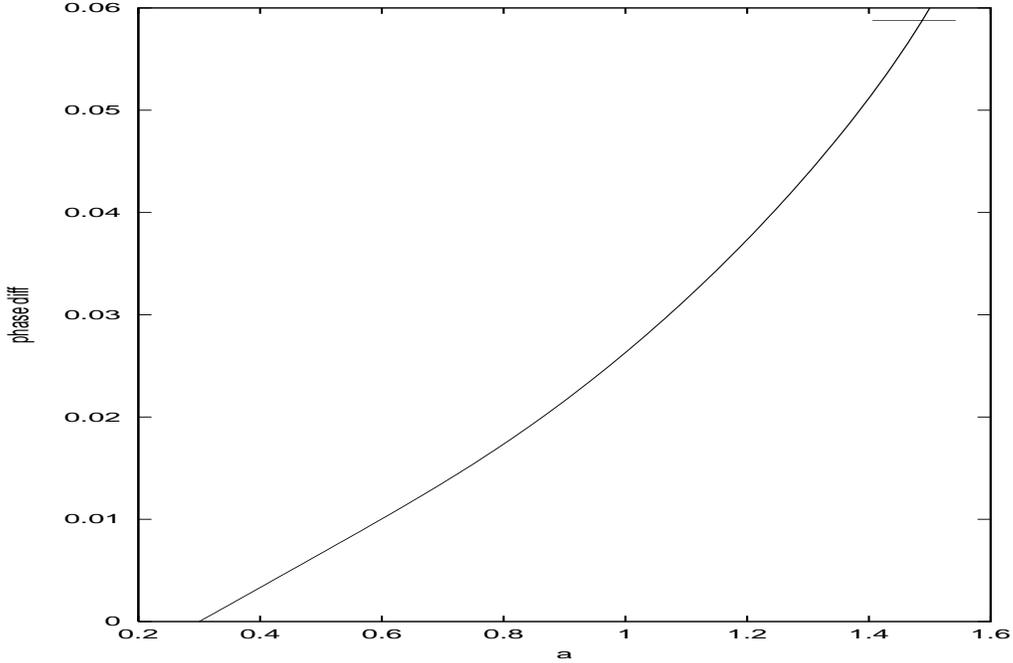,width=5.5in,height=3.5in} 
\caption{Plot of phase difference, between the solution of linear and 
nonlinear EOM of simple pendulum, after the first oscillation. $\omega 
= 4$ } 
\label{fig:2}
\end{center} 
\end{figure} 

In figure (\ref{fig:1}) we have plotted the displacement at various
times with the initial displacement. In figure(\ref{fig:2}) we have
plotted the phase difference introduced between the two displacements
which we get by solving the linear (SHM) eqn.(\ref{eq:3}) and the
non-linear eq.(\ref{eq:1}) after 
the first oscillation w.r.t. the initial displacement. As stated,
${\rm \alpha}$ is the constant delay introduced in successive
oscillations, hence it is sufficient to plot between initial
displacement and ${\rm \alpha}$ after the first oscillation. As can be
seen ${\rm f(\alpha)}$ is essentially a function of the initial
displacement. We have also tried to fit polynomial function of f(a) to
the numerical results which we get by solving eq.(\ref{eq:1}), the
result is :
\beq
f(a) = a + b x + c x^2
\label{eq:11}
\eeq
with
$$
a = - 0.0016674238  ~,~ b = 0.0026202282 ~,~ c = 0.024899586
$$

\section*{Conclusion}
In summary we have shown the difference in the results when we use the 
linear EOM (eq.(\ref{eq:3}) and non-linear EOM (eq.(\ref{eq:4}) for 
describing a pendulum. The time taken to complete an oscillation decreases 
with successive oscillations for a pendulum whose initial displacement is 
large. In short it seems that under this condition the oscillator 
will oscillate more rapidly. Thus, in case of a pendulum oscillating under the 
non-linear EOM condition, a modified value of 
$\omega$ has to be considered which is a function of the initial displacement. 
Apart from this modification the solution of non-linear EOM can also be taken 
to be harmonic.

\vfill
\end{document}